\documentclass[sigconf]{acmart}
\usepackage{booktabs} 
\usepackage{amsthm}
\usepackage{amsmath}
\usepackage{subcaption}
\usepackage[inline]{enumitem}
\usepackage{tipa}
\usepackage[ruled]{algorithm2e} 

\renewcommand\footnotetextcopyrightpermission[1]{} %

\newtheorem{assumption}{Assumption}
\theoremstyle{definition}
\newtheorem{definition}{Definition}

\acmConference{The Web Conference 2018}{April 2018}{Lyon, France}
\begin{document}
\title{Detection of the Prodromal Phase of Bipolar Disorder from Psychological and Phonological Aspects in Social Media}

\author{Yen-Hao Huang}
\affiliation{%
  \institution{National Tsing Hua University}
  \city{Hsinchu} 
  \state{Taiwan} 
  \postcode{30013}
}
\email{yenhao0218@gmail.com}

\author{Lin-Hung Wei}
\affiliation{
  \institution{National Tsing Hua University}
  \city{Hsinchu} 
  \state{Taiwan} 
  \postcode{30013}
}
\email{adeline80916@gmail.com}

\author{Yi-Shin Chen}
\affiliation{
  \institution{National Tsing Hua University}
  \city{Hsinchu} 
  \state{Taiwan} 
  \postcode{30013}
}
\email{yishin@gmail.com}






\renewcommand{\shortauthors}{Huang et al.}

\begin{abstract}
Seven out of ten people with bipolar disorder are initially misdiagnosed and thirty percent of individuals with bipolar disorder will commit suicide. Identifying the early phases of the disorder is one of the key components for reducing the full development of the disorder.
In this study, we aim at leveraging the data from social media to design predictive models, which utilize the \textit{psychological} and \textit{phonological features}, to determine the onset period of bipolar disorder and provide insights on its prodrome.
This study makes these discoveries possible by employing a novel data collection process, coined as \textit{Time-specific Subconscious Crowdsourcing}, which helps collect a reliable dataset that supplements \textit{diagnosis information} from people suffering from bipolar disorder.
Our experimental results demonstrate that the proposed models could greatly contribute to the regular assessments of people with bipolar disorder, which is important in the primary care setting.

\end{abstract}

%
%


\maketitle

\section{Introduction}

Bipolar disorder (BD) is a common mental illness characterized by recurrent episodes of mania/hypomania and depression, which is found among all ages, races, ethnic groups and social classes. 
The regular assessment of people with BD is an important part of its treatment, though it may be very time-consuming~\cite{sit2004women}.
There are many beneficial treatments for the patients, particularly for delaying relapses.
The identification of early symptoms is significant for allowing early intervention and reducing the multiple adverse consequences of a full-blown episode.
Despite the importance of the detection of prodromal symptoms, there are very few studies that have actually examined the ability of relatives to detect these symptoms in BD patients.~\cite{sierra2007prodromal} 
For the purpose of early treatment, the challenge leads to: \textbf{how to identify the prodrome period of BD}.
Current studies are thus aimed at detecting prodromes and analyzing the prodromal symptoms of manic recurrence in clinics.

With regards to the symptom of social isolation, people are increasingly turning to popular social media, such as Facebook and Twitter, to share their illness experiences or seek advice from others with similar mental health conditions.
As the information is being shared in public, people are subconsciously providing rich contents about their states of mind.
In this paper, we refer to this sharing and data collection as \textit{time-specific subconscious crowdsourcing}.

In this study, we carefully look at patients who have been diagnosed with BD and who explicitly indicate the diagnosis and time of diagnosis on Twitter.
Our goal is to both predict whether BD rises on a given period of time, and to discover the prodromal period for BD. 
It's important to clarify that our goal doesn't seek to offer a diagnosis but rather to make a prediction of which users are likely to be suffering
from the BD. The main contributions of our work are:
\begin{itemize}
  \item Introducing the concept of time-specific subconscious crowdsourcing, which can aid in locating the social network behavior data of BD patients with the corresponding time of diagnosis.
  \item A BD assessment mechanism that differentiates between prodromal symptoms and acute symptoms.
  \item Introducing the phonological features into the assessment mechanism, which allows for the possibility to assess patients through text only.
  \item An automatic recognition approach that detects the possible prodromal period for BD.
\end{itemize}

\section{Related Work}

Social media resources have been widely utilized by researchers to study mental health issues.
The following literature emphasizes on data collection and feature engineering, including subject recruitment, manual data collection, data collection applications, keyword matching, and combined approaches. 
The clinical approach for mental disorders and prodrome studies are also discussed in this section.

\textbf{Subject recruitment}: Based on customized questionnaires and contact with subjects, Park et al.~\cite{park2012depressive} recruited participants for the Center for Epidemiologic Studies Depression scale(CES-D)~\cite{radloff1977ces} and provided their Twitter data.
By analyzing the information contained in tweets, participants were divided into normal and depressive groups based on their scores on CES-D.
An approach like this one requires expensive costs to acquire data and conduct the questionnaire.

\textbf{Manual and automatic data collecting}: Moreno 
et al.~\cite{moreno2011feeling} collected data via the Facebook profiles of college students reviewed by two investigators. They aimed at revealing the relationship between demographic factors and depression.
Similarly, in our work, we invest on manual efforts to collect and properly annotate our dataset.
In addition, there are many applications built on top of social networks that provide free services where users may need to input their credentials and profile information in exchange for interesting analytics or insights.
These applications need to be well-designed to attract users, otherwise it is difficult to focus on a specific type of user. 

\textbf{Keywords matching}: Coppersmith et al.~\cite{coppersmith2014quantifying} made use of regular expressions in social media posts.
For instance, a user who states ``I was diagnosed with X'' is identified as a patient who has mental illness X. 
With this approach, people diagnosed with different illnesses are collected separately.
Although this keyword matching provides a higher precision with regards to detecting true positives, the strict filter makes the number of matched users very small.
In our approach, we also employ keyword matching, but the keywords are more flexible.
Additionally, Coppersmith et al.~\cite{coppersmith2014quantifying,harman2014measuring,coppersmith2015adhd} focused on linguistic features and patterns of life features (specific for social network behaviors) to build a predictive model for mental disorders.

\textbf{Combined approach}: Some previous studies have combined the above approaches to accumulate desirable data. De Choudhury~\cite{de2013predicting} used a two-stage approach to collect Twitter data from mothers of newborns. 
First, the mothers were selected by matching keywords related to birth announcements in tweets.
Second, the manual verification task was delegated to Amazon's Mechanical Turk to improve the reliability of data.
For postpartum depression, a statistical model was built by using linguistic features, a social graph, and engagement in social networking websites. 

\textbf{Subconscious crowdsourcing}: Chang et al.~\cite{chang2016subconscious} focused on collecting high-quantity data. 
Their approach combines manual and automatic efforts by manually filtering possible mentally ill users from patient participation groups and user profiles on Twitter.
With this approach, highly self-aware users are selected, which is not one of our objectives.
Inspired by Chang et al.s' approach, however, we take the advantage of keyword matching to maximize the resources found on Twitter.

\textbf{Clinical approach}: For the study of prodromal features there have been several \emph{retrospective works}~\cite{howes2011comprehensive, berk2007history} that have found that the majority of BD patients experience symptoms such as episodic mood changes, irritability, or impulsivity before the onset of the first episode of BD. 
Sahoo et al.~\cite{sahoo2012detection} tracked the recurrence of patients, proceeded with the self-diagnosis of DSM-IV (The Diagnostic and Statistical Manual of Mental Disorders, Fourth Edition), and pointed out the difficulty of patients' self-awareness.
Meter et al.~\cite{van2016bipolar} also surveyed prodromes and underlined the importance of a sleeping monitor, physical activity, and family attention.

The previously mentioned findings certainly support the existence of a prodromal phase.
Nevertheless, these studies may have been influenced by a recall bias. 
Furthermore, most clinical studies mainly focus on patients found in hospitals, which cost substantially and leave the problem of late diagnosis unsolved.

Among the aforementioned researches on social media, none focus on the detection of the acute and prodromal phases of a mental disorder. 
Clinical researchers have dived into the details of prodromes, providing many traceable features. 
However, for some recurrent mental disorders, it is difficult to closely monitor the mental status of a patient in person after they leave the clinic, even though regular assessment is important in a primary care setting. 
In our work, we commit to seeking resources on social media platforms and propose an approach using linguistics, behavioral, and clinical features.

\section{Methodology}


\subsection{Overviews}


The objective of this paper is to recognize the prodromal phases of bipolar disorder by linguistic and phonological features.
Given a user at a specific time as the input, the recognition system can perform a regular assessment of BD.
The problem is formally defined as follows:

\begin{definition}[Problem Statement]
Given a specific time $s$, a time frame $\alpha$, and a user $k$, we can collect a set of tweets $\Gamma^{(k)}_{s,\alpha} = \{ \omega^{(k)}_t \mid t \in  [s- \alpha, s]\}$ and derive the model $f$ with the objective to map the set $\Gamma^{(k)}_{s,\alpha}$ to a BD onset probability $l^{(k)}_{s,\alpha} = f(\Gamma^{(k)}_{s,\alpha})$, which is a number in $[0,1]$, where $0$ denotes no onset of BD and $1$ denotes a full onset.
\label{def:problemstatement}
\end{definition}



Since BD is a recurrent mental illness and the onset symptoms are intermittent on a monthly basis, identifying its possible prodromal phases from social media data becomes a challenging task. 
Our solution introduces the \textit{time of diagnosis} of the illness to our recognition model.
The time of diagnosis specifies the exact date when the person was diagnosed as a patient with BD by a certified psychiatrist.
Based on the discussion above, assumption ~\ref{assumption:arising} is made.

\begin{assumption}
When people are diagnosed with BD, the signs and symptoms are observable.
\label{assumption:arising}
\end{assumption}

Our proposed prodromal recognition method is called \textit{Illness Behavior Inference}.
Its core contribution is the \textit{BD onset predictive model(BDOPM)}, a supervised learning process that predicts the BD onset probability given the following information: time $s$, user $k$, tweets $\Gamma^{(k)}_{s,\alpha}$, time frame $\alpha$, and is optimized by the time of diagnosis $\tau^{(k)}$.  



Not only is this predictive model able to capture the acute phases of bipolar disorder, but can also recognize the different behaviors for a BD patient over time given a \textit{time of diagnosis} indicator $\tau$.
The construction of the BD onset predictive model is divided into five stages:
\begin{enumerate}
\item Time-specific Subconscious Crowdsourcing
\item Time-domain Modeling
\item Psychological Feature Extraction
\item Phonological Feature Extraction
\item Model Construction and Evaluation
\end{enumerate}

\subsection{Time-specific Subconscious Crowdsourcing}

As social media reflect a mirror aspect of real life, people widely share their feelings, moods, and opinions on them.
The subconscious crowdsourcing~\cite{chang2016subconscious}, as implied by the term \textit{subconscious}, indicates that people are unaware and subconsciously providing information, including their mental status.
Users with mental illnesses share their common experiences, such as mania behavior or relapses of an illness.
This behavior represents the concept of the term \textit{crowdsourcing}.
We can reduce the cost of data collection by accessing information that is richly stored in social media platforms.

Since our task is to cope with the recurrent characteristic of BD, every user cannot be regarded as an onset patient with BD the whole time.
In this stage, we provide four steps for retrieving and collecting a list of BD patients and their \textit{time of diagnosis} $\tau$ by combining automatic matching with manual efforts.

\subsubsection{\textbf{Target Seeking}} 
To efficiently retrieve the targeting dataset without losing too much information, we perform an automatic process---a \textit{minimum keyword search}---on the platform to obtain related information, followed by labeling with the search keys.
The \textit{minimum keyword search} procedure is formally defined below and utilized to collect the BD user data.

\begin{definition}[Minimum Keyword Search]
Let Twitter be composed of a set of tweets $\Gamma = \{\omega\}$ and tweet $\omega$ be composed of a set of $n$-grams $\mathbb{V}^{n=1}_\omega = \{v_\omega \mid v_\omega \in \omega\}$. 
We then select a set of \textit{unigrams} $\Phi$ as filter keys.
Given the search function $g(\Gamma, \Phi)$, it returns a set of candidate tweets $\Gamma_\Phi = \{\omega \mid \Phi\ \subseteq \mathbb{V}^{n=1}_\omega \}$.
\end{definition}
To search for users with BD, the keywords~\textbf{diagnosed} and \textbf{bipolar} are utilized to perform the minimum search on Twitter.
Twitter users who have mentioned these keywords in at least one tweet will be selected as users with BD and the corresponding tweet(s) will be collected as diagnosis tweet statements.
Different to Coppersmith et al.'s keyword approach~\cite{coppersmith2014quantifying}, our minimum search provides more flexible filtering, which leads to more tweet matches and more BD candidates on Twitter.

\subsubsection{\textbf{Time of Diagnosis Identification}} After the previous step is completed, we have collected many possible BD users.
This dataset of users and tweets, however, may contain a lot of noise. 
An additional filtering step is thus required. 
In our case, which is to target the onset periods, we further modify the \textit{minimum keyword search} in order to retrieve every time-related diagnosis tweet $\Gamma_\Phi = \{\omega \mid (\Phi\ \cap \mathbb{V}_\omega) \neq \emptyset \}$ with time-related keywords, such as \textit{today}, \textit{last}, \textit{months}, etc.

A manual step is required here to estimate the time of diagnosis $\tau$ more precisely.
For example, with the keywords ``diagnosed'' and ``bipolar'', we may get a tweet like ``I~was diagnosed Bipolar Disorder last year this month''. In this example, we are able to obtain not only the BD user but also their time of diagnosis.
The time of diagnosis confirms that the user is a real BD user, indicates a more precise onset period, and reduces the chance to extract wrong features by processing the tweets in the non-onset period.

To make the diagnosis more precisely, the diagnosis tweets are filtered manually by the criteria that if it is able to recognize that the twitter user are diagnosed on which year and month. In this case, the unclear diagnosis tweets such as ``I~was diagnosed Bipolar Disorder last year'' or ``I~was diagnosed Bipolar Disorder few months ago'' are deleted, because their diagnosed time are not clear enough.

The tweet author $k$ of a qualified tweet is following extracted, in which the $\tau$ of the tweet can be estimated.
The output of this step is a set of qualified BD users $\mathbb{K}^{(b)}$ associated with their time of diagnosis $\tau^{(k)}$.

\subsubsection{\textbf{Tweet Re-collection}} We download all tweets $\Gamma^{(k)},$ $k \in \mathbb{K}^b$ from Twitter. 
In our case, we also retrieve each user's timezone and location information if available.

\subsubsection{\textbf{Language Filter and Active User Filter}}
Our method focuses on English content, for which reason any user who has more than 50 percent of their posts containing hyperlinks or other languages are removed.
Additionally, only users with more than 100 tweets are included, as they are assumed to be more active users.\\

In order to optimize the \textit{BDOPM} stage, the complement dataset (non-BD users) needs to be collected.
The user IDs of Twitter are randomly sampled using the Twitter Streaming API for several months of data.
These randomly-sampled users are labled as regular users $\mathbb{K}^{(r)}$ (non-BD) and their corresponding time of diagnosis is set to $\tau^{(k)} = 0, k \in \mathbb{K}^{(r)}$.
Subsequently, the tweets of each selected ID are downloaded, and denoted as $\Gamma^{(k)} = 0, k \in \mathbb{K}^{(r)}$.

\subsection{Illness Period Modeling}
For the purpose of onset detection, and owing to the recurrent nature of BD, we aim to find the most likely periods where BD is onset by partitioning the tweets into time periods. 
Based on Assumption~\ref{assumption:arising}, the $\tau$ and the length of time $\alpha$ are applied to identify the possible acute phase.
The illness's acute phase is formally defined as:
\begin{definition}[Bipolar Disorder Onset Period]
The acute phase is given as $ \mathbb{S}^\tau_\alpha = \{t \mid t \in [\tau-\alpha, \tau ]\}$.
\end{definition}

\begin{figure}[h!]
    \centerline{\includegraphics[width=\linewidth]{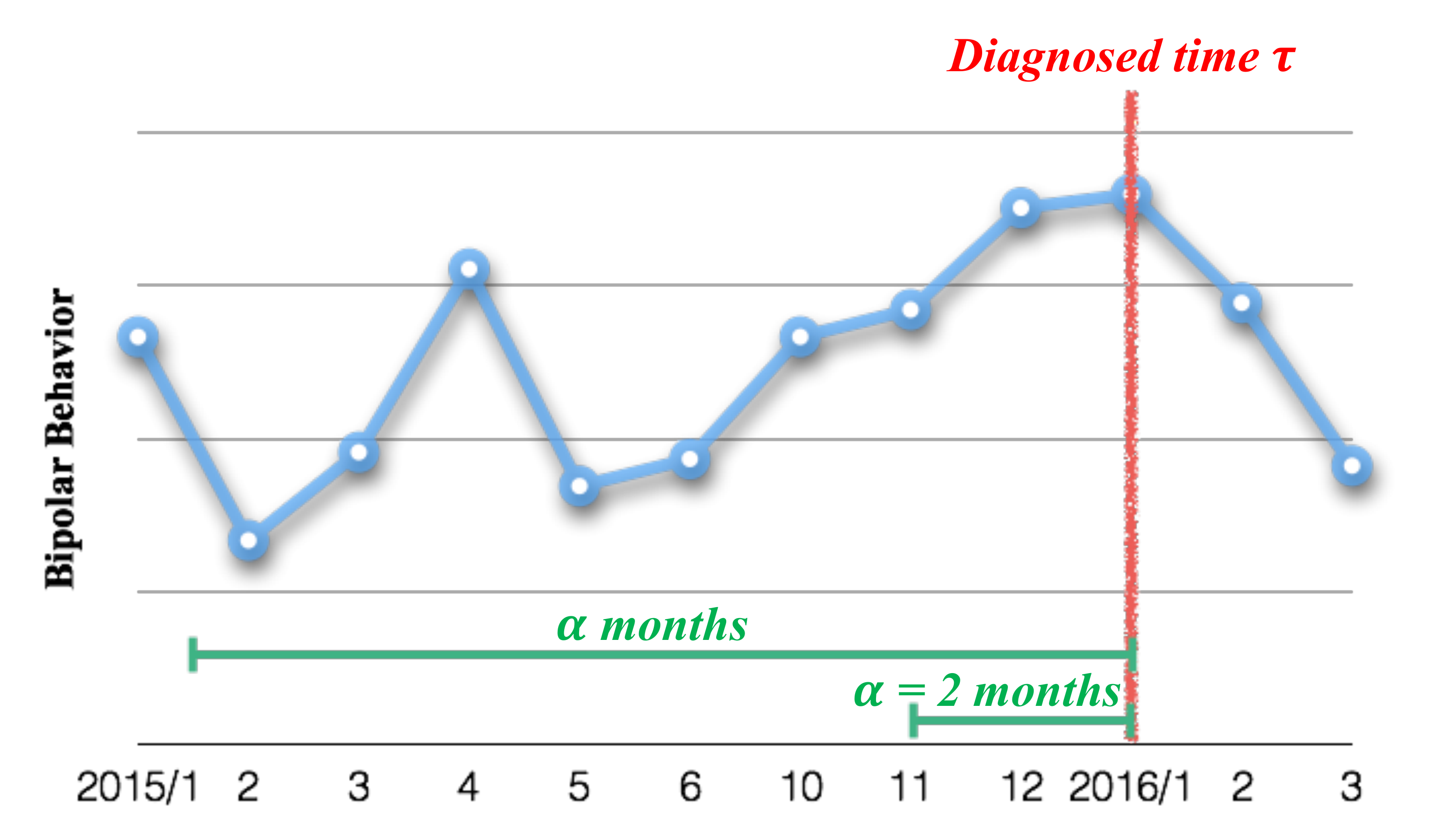}}
    \caption{Illness Period Modeling}
    \label{fig:time_seg}
\end{figure}

 For demonstration purposes, the concept is illustrated in Figure~\ref{fig:time_seg}, assuming a BD onset period of two months, (i.e., $\alpha = 2$).
 After the possible onset period $\alpha$ is defined, the user's tweets are extracted as $\Gamma^{(k)}_{\tau, \alpha} = \{\omega_t \mid t \in \mathbb{S}^\tau_\alpha\}$ and are considered as the BD onset training data.
 For non-BD users, the tweets during period $\alpha$ are treated as the training data for regular users.
 That is, the specific time $s$ in Definition~\ref{def:problemstatement} is set to be the $\tau$ for the model-training step for the best approximation to the BD onset behavior.


\subsection{Psychological Feature Extraction}
In order to recognize a case of BD, the \textit{BDOPM} is customized to detect the following clinical exhibitions:

\begin{enumerate*}
    \item Rapid cycling between moods.
    \item High energy, over talking, and tendency towards anger.
    \item Lessened need for sleep.
    \item Cyclical depression or mania.
\end{enumerate*}

As discussed in the related work, there are various types of linguistic and behavioral features that are important for BD onset detection.
In this step, two main types of psychological features are introduced: 
\begin{enumerate*}
    \item \textit{Word-level} features and \textit{BD Pattern of Life} features.
\end{enumerate*}

\subsubsection{\textbf{Word-level Features}}
With respect to the linguistic features for BD, the \textit{Character n-gram language features(CLF)} and \textit{LIWC} metrics are designed to capture it.
The \textit{CLF} utilizes $n$-grams to measure the comment words or phrases used by users. 
The \textit{tf-idf} is utilized in our score-calculating method, the $tf$ is the frequency of an $n$-gram and the document $d$ of $df$ is defined as each particular twitter user $k$. 
The formula for the \textit{tf-idf} is thus given as:

\begin{equation}
tfidf^{(k,\tau,\alpha)}_{v^n} = freq^{(k,\tau,\alpha)}_{v^n} \times \log \frac{K}{1 + freq^{(K,\tau,\alpha)}_{v^n}}
\end{equation}
The $freq^{(k)}_{v^n}$ is the frequency of n-gram $v^n$, which is $n \in \{1,2\}$ for a specific user~$k$.
The $K$ denotes the total number of users in the dataset, and $freq^{(K,\alpha)}_{v^n}$ denotes the number of user whose tweets contain the terms of $v^n$. 
The resulting \textit{tf-idf} vectors are then normalized by the Euclidean norm:
\begin{equation}
{tfidf^{(k,\tau,\alpha)}_{v^n}}_{norm} = \frac{tfidf^{(k,\tau,\alpha)}_{v^n}}{\sqrt{(tfidf^{(1,\tau,\alpha)}_{v^n})^2, \ldots, (tfidf^{(K,\tau,\alpha)}_{v^n})^2}}
\end{equation}

The \textit{LIWC} features are designed to capture the psychological terms frequently used by a BD user.
The words that are related to psychological terms (e.g., emotion, affection, and depression) are selected to represent the features.
Based on the dictionary obtained from the Linguistic Inquiry and Word Count (LIWC)~\cite{pennebaker2007liwc2007} lexicon, we then calculated the ratio of each category (64 in total) for each user.
The LIWC score of category $c_i$ for a given user $k$ is defined as follows:
\begin{equation}
LIWC^{(k,\tau,\alpha)}_{c_i} = \frac{freq_{c_i, \Gamma^{(k)}_{\tau,\alpha}}}{freq_{\Gamma^{(k)}_{\tau,\alpha}}}
\end{equation}
The LIWC scores are finally normalized by the total number of user tweets $|\Gamma^{(k)}_{\tau,\alpha}|$.

\subsubsection{\textbf{Bipolar Disorder Pattern of Life Features (BDPLF)}}
To capture the features of the BD onset period in the psychological aspect, we propose customized \textit{BD pattern of life features (BDPLF)}, inferred from~\cite{chang2016subconscious, coppersmith2014quantifying}.
The \textit{PLF} is designed to represent psychological features, such as emotional patterns and the behavioral tendency of users by measuring polarity, emotion, and social interactions.
To construct the full \textit{BDPLF}, there are five categories:

\begin{itemize}
  \item \textbf{Age and Gender}:
  Sit et al.~\cite{sit2004women} studied the gender effects on BD, indicating that women with BD are more likely to have Bipolar Disorder(Type II) symptoms than men.
  We make use of the age and gender predictor proposed by Sap et al.~\cite{sap2014developing}, which is based on lexica in social media.
\item \textbf{Mood Polarity Features}:
 Owing to the fact that BD patients experience \textit{rapid mood changes}, \textit{sentiment analysis} is firstly adapted to obtain the sentiment polarity portrayed by each user's tweets $\Gamma^{(k)}$.
 To obtain the sentiment of tweets, the online tool Sentiment140 is used, based on Go et al.'s work~\cite{go2009twitter}.
 The tool classifies the contents of tweets into three polarity categories: \textit{positive}, \textit{negative}, and \textit{neutral}.
 We further partition those three categories into five different sub-features: \textit{positive ratio}, \textit{negative ratio}, \textit{positive combo}, \textit{negative combo}, and \textit{flips ratio}.
  \item \textbf{Emotional Scores}:
  Beyond the sentiments, an emotion detection tool proposed by Argueta et al.~\cite{argueta2015unsupervised} is employed to classify the tweets into eight emotion categories: \textit{joy},
  \textit{surprise}, \textit{anticipation}, \textit{trust}, \textit{sadness}, \textit{disgust}, \textit{anger}, and \textit{fear}.
  The emotion classification results are further transformed into emotion scores as follows:
  \begin{equation}
  es_{{e_i},\Gamma^{(k)}_{\tau,\alpha}} = \frac{e_{i,\Gamma^{(k)}_{\tau,\alpha}}}{e_{count}}
  \end{equation}
  where $e_i$ is the $i$th emotion and $e_{count}$ is the total number of emotions. 
  
  \item \textbf{Social Features}:
  Social features are designed to capture a user's interaction with other users on the social media platform and how frequently they engage on Twitter.
  There are five designed social features, of which the basic four are \textit{tweeting frequency, mention ratio, frequent mentions}, and \textit{unique mentions}.
  \item \textbf{Insomnia and Over-Talking Features}:
   Our approach considers two additional social features to capture the \textit{sleep disturbance} and \textit{over-talking} symptoms: 
  \begin{itemize}
     \item \textbf{Late Tweet Frequency}
     The less-sleep symptom is a prominent characteristic of BD.
     In order to capture this feature, our approach performs a \textit{timezone conversion} step to convert UTC time to each user's local time based on their timezone and location to make sure the user's tweeting time is recorded correctly.
     This feature is determined by the daily frequency of posts that a user posts between midnight and 6:00 am in their local time.
     \item \textbf{Tweet Rate Difference}
     For the over-talking feature, this feature is designed to capture the difference in tweeting behavior from individuals.
     First, we segment each user's tweets based on a \textit{slide window approach}.
     Next, for each segmentation period, we calculate the average tweet frequency per day.
     The maximum tweeting rate difference is the max difference of the tweet frequency for each given user.
  \end{itemize}
\end{itemize}
From the features listed above, the \textit{bipolar disorder pattern of life feature} scores $BDPLFs$ are formally illustrated as follows:
\begin{equation}
    BDPLFs^{(k,\tau,\alpha)}_{\xi_m} = BDPLF_{\xi_m}(\Gamma^{(k)}_{\tau,\alpha})
\end{equation}
where $\xi_m$ denotes each sub-feature in \textit{BDPLF}, and $BDPLF_{\xi_m}$ denotes the sub-functions for the sub-features, which return the corresponding feature scores.

\subsection{\textbf{Phonological Feature Extraction}}
For phonological features, we consider that words have their own energy.
People in different states have distinct tendencies to use words.
Generally, people who use more high-energy words are usually in states of excitement or anger. 
Contrastingly, people who are in states of powerlessness or helplessness use more low-energy words.
People who suffer from depression show powerlessness more often than ordinary people do.

As a BD patient is encountered with mania or hypermania symptoms, they will tend to reveal high energy, talk more, and have a tendency for anger.
When the illness is onset, the expressions of a BD patient will be especially influenced by these symptoms and leave emotions in the expression interface, such as in tweets.
The \textit{Energy of Words} feature is thus designed to capture the mania expression based on  Assumption~\ref{assumption:energy_of_words}. 

\begin{assumption}
Writing is accompanied by reading in the brain.
Words carry their own energy, that is, when people try to write a word, people simultaneously read that word in their brains.
The energy of a person can therefore be inferred by the statement at the moment they were writing.
\label{assumption:energy_of_words}
\end{assumption}

Formally, we define the key concept of how the energy of words are measured as follows:
\begin{definition}[Energy Score] 
The \textit{Energy Score} $es_v$ is the energy of a word.
It can be expressed as $es_v = ES(v)$, where $v$ denotes a specific word and $ES$ denotes the \textit{Energy Score Function}.
\label{def:energy_score}
\end{definition}

\begin{definition}[Energy Score Function] 
The Energy Score Function $ES(v) = PF(IPA(v))$, where $IPA(v)$ represents the function that converts a word $v$ to its IPA form.
$PF$ denotes the function that returns phonological feature scores in the IPA form.
    \label{def:energy_score2}
\end{definition}


In the following sections, we give details to clarify the entire process of how to quantify the Energy Score step by step in Definition~\ref{def:energy_score2}.

\subsubsection{\textbf{Pronunciation Standardization}}
\begin{figure*}
    \centering
    \begin{subfigure}[b]{0.6\textwidth}
            \includegraphics[width=\linewidth]{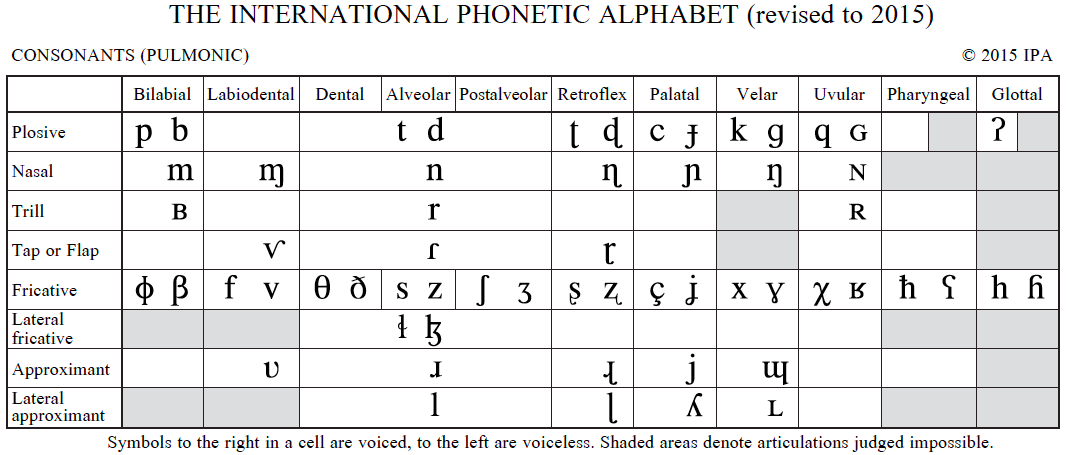}
            \caption{IPA Consonants Table}
            \label{fig:ipaTableConsonants}
    \end{subfigure}
    \centering
    \begin{subfigure}[b]{0.33\textwidth}
            \includegraphics[width=\linewidth]{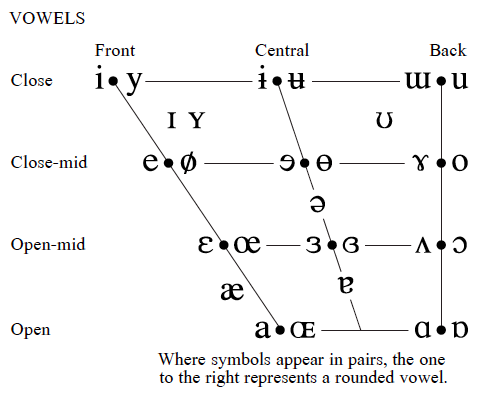}
            \caption{IPA Vowels Table}
            \label{fig:ipaTableVowels}
    \end{subfigure}
    \caption{The International Phonetic Alphabet (IPA)\cite{IPA2015IPA}, an alphabetic system of phonetic notation based primarily on the Latin alphabet. It was devised by the International Phonetic Association as a standardized representation of the sounds of spoken language.}\label{fig:ipaTable}
\end{figure*}

To standardize the pronunciation of each alphabet $a_i$, a standardization method denoted \textit{IPA Standardization Function} $IPA(v)$, shown in Definition~\ref{def:energy_score2}, is performed to transfer English words into their IPA (International Phonetic Alphabet) form\cite{IPA2015IPA}.
The table of IPA phonetic transcriptions of consonants and vowels is shown in Figures~\ref{fig:ipaTableConsonants} and~\ref{fig:ipaTableVowels}.
Each IPA symbol is a \textbf{phoneme}, which is one of the units of sound that distinguish one word from another.
Hence, the \textit{IPA standardization function} is defined below to transfer a word to its transcription.
\begin{definition}[IPA Standardization Function]
Let $\mathbb{P}$ be a set of phonemes and $p_j$ be a phoneme, where $p_j \in \mathbb{P}$.
The IPA of a word $v$ is denoted as $\mathbb{P}_v = IPA(v)$, where $\mathbb{P}_v = \{p_1, ..., p_J\}$.
    \label{def:ipa_transform}
\end{definition}

A word goes through the \textit{IPA Transcription Function}, which outputs its transcription consisting of a sequence of phonemes.
For instance, given the word $v = $ ``\textit{folder}'', it is able to retrieve the output $\mathbb{P}_v = $ through $IPA(v)$, where $\mathbb{P}_v =$ \{\textipa{f, oU, l, d, @, r}\}.

\subsubsection{\textbf{Phonological Feature Score (PF Score)}}
In this subsection, the phonological features of phonemes are utilized to define how much energy a word carried.
There are two steps for the \textit{Phonological Feature Score Function} to be completed:
    

First, let's define the energy score of each phoneme $p_j$.
From the existing phonological feature systems, the eSPE (extended Sound Pattern of English)~\cite{cervnak2017speech} is introduced to describe the phonological features of phonemes.
The eSPE consists of 21 binary features.
It describes what kinds of articulators and which part of positions are used during speaking.
In our work, 19 features from the eSPE are chosen, with the exception of ``voiced'' and ``silence''.
The eSPE shows that each phoneme corresponds to some phonological feature, such as vowel, fricative, nasal and so on.
The notations of positive and negative represent independently if a phoneme has the corresponding feature or not.
For instance, the phoneme ``\textipa{oU}'' has phonological features \textit{vowel}, \textit{high}, \textit{mid}, \textit{back}, \textit{continuant}, \textit{round}, and \textit{tense}.

Instead of using binary features directly, these relative phonological features are grouped into the following 8 categories based on characteristics of phonology.
Additionally, the corresponding scores are assigned according to their \textit{difficulty of pronunciation}:
\begin{itemize}
    \item \textbf{Oral Cavity ($OC$):} the phonological features of anterior, back and approxim., with scores $ds_{OC1}$, $ds_{OC2}$, $ds_{OC3}$. 
    \item \textbf{Mouth Openness ($MO$):} the phonological features of high, mid and low, with scores $ds_{MO1}$, $ds_{MO2}$, $ds_{MO3}$.
    \item \textbf{Obstruent ($Obs$):} the phonological features of continuant, labial, fricative and stop, with scores $ds_{Obs1}$, $ds_{Obs2}$, $ds_{Obs3}$.
    \item \textbf{Tongue Position ($TP$):} the phonological features of coronal, dental and retroflex, with scores $ds_{TP1}$, $ds_{TP2}$, $ds_{TP3}$.
    \item \textbf{Resonance ($Res$):} the phonological features of nasal, glottal and velar, with scores $ds_{Res1}$, $ds_{Res2}$, $ds_{Res3}$.
    \item \textbf{Vowel ($Vow$):} the phonological feature of vowel itself with the score $ds_{Vow1}$.
    \item \textbf{Round ($Rou$):} the phonological feature of round itself with the score $ds_{Rou1}$.
    \item \textbf{Tense ($Ten$):} the phonological feature of tense itself with the score $ds_{Ten1}$.
\end{itemize}
The total scores of each category are denoted as $ds_{OC}$,  $ds_{MO}$, $ds_{Obs}$, $ds_{TP}$, $ds_{Res}$, $ds_{Vow}$, $ds_{Rou}$ and $ds_{Ten}$, by summing up the scores of their sub-features. 
Based on these categories, the phonological score for each phoneme is vectorized as :
\[
pf_{p_j} = | \begin{matrix} ds_1 & ds_2 & \ldots & ds_8 \end{matrix} |
\]
where each element corresponds to the scores of each category. 
To calculate word energy, the \textit{Phonological Feature Score Function} is formulated as follows:
\begin{equation}
PF(\mathbb{P}_v) = \sum_{j=1}^{J}pf_{p_j}
\label{equation:pf_scores}
\end{equation}
where each element of $PF(P_v)$ denotes the corresponding scores of each category. 
The \textit{energy of a user} is finally represented as:
\begin{equation}
      eu^{(k)}_{\tau,\alpha} = \frac{\sum es^{(k,\tau,\alpha)}_v}{v_{count}} = \frac{ES( v | v \in {\mathbb{V}^{n=1}_{\tau,\alpha}})}{|\mathbb{V}^{n=1}_{\tau,\alpha}|}
\end{equation}
where $\mathbb{V}^{n=1}_{\tau,\alpha}$ denotes a set of single words decomposed from all the tweets in $\Gamma^{(k)}_{\tau, \alpha}$. 
The energy $eu$ of a user is final normalized using the total frequency of words $v_{count}$.

\subsection{Model Training and Evaluation}
In this work, an ensemble supervised machine learning approach, \textit{Random Forest Classifier}, has been empirically chosen to be our training model $\hat f$, as it is resistant to overfitting and is able to perform both classification and regression tasks.

With the feature extraction steps described previously, the BD onset probability can be represented as:
\begin{equation}
l^{(k)}_{\tau,\alpha} = f(\Gamma^{(k)}_{\tau,\alpha}) = \hat f(h(\Gamma^{(k)}_{\tau,\alpha})), 
\end{equation}
where $h$ denotes the feature extraction functions, which are \textit{word-level (CLF and LIWC)}, \textit{pattern of life}, and \textit{phonological} feature functions.
In order to compare feature performance, the \textit{BDOPM} is trained for each of the five types using different features, which are \textit{word-level(character n-gram language model(CLM) and LIWC model)}, \textit{bipolar disorder pattern of life model} and \textit{phonological model}:
\begin{equation}
    l^{(k)}_{\tau,\alpha} = 
\begin{cases}
    \hat f_{tfidf}(tfidf^{(k,\tau,\alpha)}_{v^n})\\
    \hat f_{liwc}(LIWC^{(k,\tau,\alpha)}_{c_i})\\
    \hat f_{BDPLF}(BDPLFs^{(k,\tau,\alpha)}_{\xi_m})\\
    \hat f_{eu}(eu^{(k)}_{\tau,\alpha})\\
    \hat f_{BDPLF\&eu}(BDPLFs^{(k,\tau,\alpha)}_{\xi_m} \cup eu^{(k)}_{\tau,\alpha})
\end{cases}
\end{equation}


To measure the performance of our approach, we emphasize on whether the predictive model is able to distinguish a given user is in a BD onset period or not.
To evaluate our model, we use \textit{10-fold cross validation}, and measure the model's overall precision and recall.
Additionally, a \textit{psychiatrist evaluation} is conducted in our work for qualitative analysis.

\section{Prodrome Recognition}

The definition of the prodrome stage is made below and a prodrome recognition approach is proposed to locate the possible prodromal period for BD.
The prodromal period generally refers to the time interval between the onset of the first prodromal symptom and the onset of the characteristic signs/symptoms of the fully developed illness.

In order to capture the prodromal period, our approach is able to reveal the onset probability for each user through \textit{BDOPM}. 
In this step, a week by week sliding window approach is adopted to construct a bipolar disorder onset timeline, which presents onset probabilities along time.
\begin{definition}[Sliding Window Segmentation]
Let a time segmentation sequence $\psi^{(k,\alpha)}_q = \{t \mid t \in [s^{(k)}_q-\alpha,s^{(k)}_q]\}$ be in ascending order, where $s^{(k)}_q$ is the given time for user $k$.
The objective of the segmentation algorithm is to iterate over $\psi^{(k,\alpha)}_q$ with $s^{(k)}_q$ minus \textit{1 week} for each iteration $s^{(k)}_Q, s^{(k)}_{Q-1}, \ldots, s^{(k)}_q, \ldots, s^{(k)}_1$ and output the user's time segmentation sequence $\Psi^{(k,\alpha)} = \{ \psi^{(k,\alpha)}_1, \ldots, \psi^{(k,\alpha)}_Q \}$,
\end{definition}
With a user's time segmentation set $\Psi^{(k,\alpha)}$, the onset probability $l^{(k)}_{s,\alpha}$ for each time segmentation $\psi^{(k,\alpha)}_q$ can be represented as $l^{(k,\alpha)}_{\psi_q}$.
\begin{definition}[Bipolar Disorder Onset Timeline]
The \textit{Bipolar Disorder Onset Timeline} consists of a sequence of BD onset probabilities~$\mathbb{L}^{(k,\alpha)}_\Psi = \{l^{(k,\alpha)}_{\psi_q}\}$.
\end{definition}
\begin{definition}[Prodromal Period]
The prodromal periods of a user are $\mathbb{S}^{(k)} = \{ t \mid t \in \psi^{(k,\alpha)}_q \cup \psi^{(k,\alpha)}_{q+1} \cup ,\ldots \}$, $l^{(k,\alpha)}_{\psi_q} \in [l_{lower}, l_{upper}]$, where $l_{lower}$ and $l_{upper}$ are the lower and upper bounds, respectively, for no onset and onset probabilities.
\label{def:prodrome}
\end{definition}

Given the \textit{bipolar disorder onset timeline} and  Definition~\ref{def:prodrome} above, our model then infers the prodromal period through Algorithm~\ref{alg:generator}.
\begin{algorithm}
\caption{Prodromal Period Locating}
\label{alg:generator}
\SetKwProg{generate}{Function \emph{generate}}{}{end}
\textbf{INPUT} $\mathbb{L}^{(k,\alpha)}_\Psi$ and $\Psi^{(k,\alpha)}$\;
qualifiedProdrome = new Array() \;
candidateProdrome = $\emptyset$ \;
$l_{lower}$ = not onset probability \;
$l_{upper}$ = onset probability \;
\ForAll{BD onset probability $l^{(k,\alpha)}_\psi$ in $\mathbb{L}^{(k,\alpha)}_\Psi$}{
  \uIf{ $l^{(k,\alpha)}_\psi$ between $l_{lower}$ and $l_{upper}$ }{
    candidateProdrome = candidateProdrome $\cup$\ $\psi$ \;
  }
  \uElseIf{ $l^{(k,\alpha)}_\psi$ > $l_{upper}$ and Length( candidateProdrome) > 0}{
    qualifiedProdrome append candidateProdrome \;
    candidateProdrome = $\emptyset$ \;
  }
 }
\textbf{RETURN} qualifiedProdrome \;
\end{algorithm}


\section{Experiments}
\subsection{Onset Prediction Performance}
\subsubsection{\textbf{Experimental Setup}}

Time-specific subconscious crowdsourcing is applied to collect the dataset.
About 10,000 tweet statements were collected from Twitter and then filtered by the keywords ``diagnosed'' and ``bipolar'' from Oct., 2006 to Dec., 2016. 
Owing to the fact that the time of diagnosis could be expressed in various ways besides natural language processing approaches, it was manually labeled when mentioned in tweets.
Moreover, self-diagnosed users were removed to ensure reliability.
After applying the previous filter, 406 users remained.
In order to calibrate the tweet time to the user's local time, the users' timezones and locations, which are public on Twitter, were also collected.
Based on the time of diagnosis, we collected user tweets between the time of diagnosis and one year before.
These tweets became our positive class in the training process. 
Random samples of users were obtained using the Twitter Streaming API and REST API, which then became our negative class in the final training dataset.

For illness period modeling, the time frame is set to be $\alpha \in \{2,3,6,9,12\}$ months.
A dataset is modeled for each $\alpha$, and active user filtering is applied in this step.
The dataset for model evaluation is summarized in Table~\ref{table:dataset}.

\begin{table}[h]
    \centering
    \begin{tabular}{c|c|c|c} 
         \hline
         Groups & Users & Tweets & Average tweets \\ [0.3ex] 
         \hline
         Regular (2 months)  & 260 &  67485 & 259 \\
         Bipolar (2 months)  & 133 & 136837 & 1028 \\ 
         Regular (3 months)  & 257 &  89141 & 346 \\
         Bipolar (3 months)  & 135 & 226460 & 1677 \\ 
         Regular (6 months)  & 267 & 136481 & 511 \\
         Bipolar (6 months)  & 124 & 353559 & 2851 \\ 
         Regular (9 months)  & 289 & 170380 & 589 \\
         Bipolar (9 months)  & 117 & 481392 & 4114 \\ 
         Regular (12 months) & 295 & 207008 & 701 \\
         Bipolar (12 months) & 111 & 586656 & 5285 \\ 
         \hline
    \end{tabular}
    \caption{The total number of accounts and tweets for different groups of users and the group of users after illness period modeling.}
    \label{table:dataset}
\end{table}

The data collection approach, \textit{time-specific subconscious crowdsourcing(TSC)}, is first compared with the subconscious crowdsourcing(SC) (Chang et al. ~\cite{chang2016subconscious} 2016).
For model performance evaluation, we compare \textit{BDOPM} with 3 baselines :
\begin{enumerate*}
    \item \textit{PLF approach} (Chang et al.~\cite{chang2016subconscious} 2016) which consisted of Age and Gender(AG), Social(Soc),  Emotion(Emot), and Polarity(Pol) features
    \item \textit{CLM approach}(Coppersmith et al.\cite{coppersmith2014quantifying} 2014, Chang et al.~\cite{chang2016subconscious} 2016)
    \item \textit{LIWC approach} (Coppersmith et al.\cite{harman2014measuring,coppersmith2014quantifying} 2014).
\end{enumerate*}

\subsubsection{\textbf{Dataset Collection Approach Comparison}}
\begin{figure}[h]
    \centerline{\includegraphics[width=0.8\linewidth]{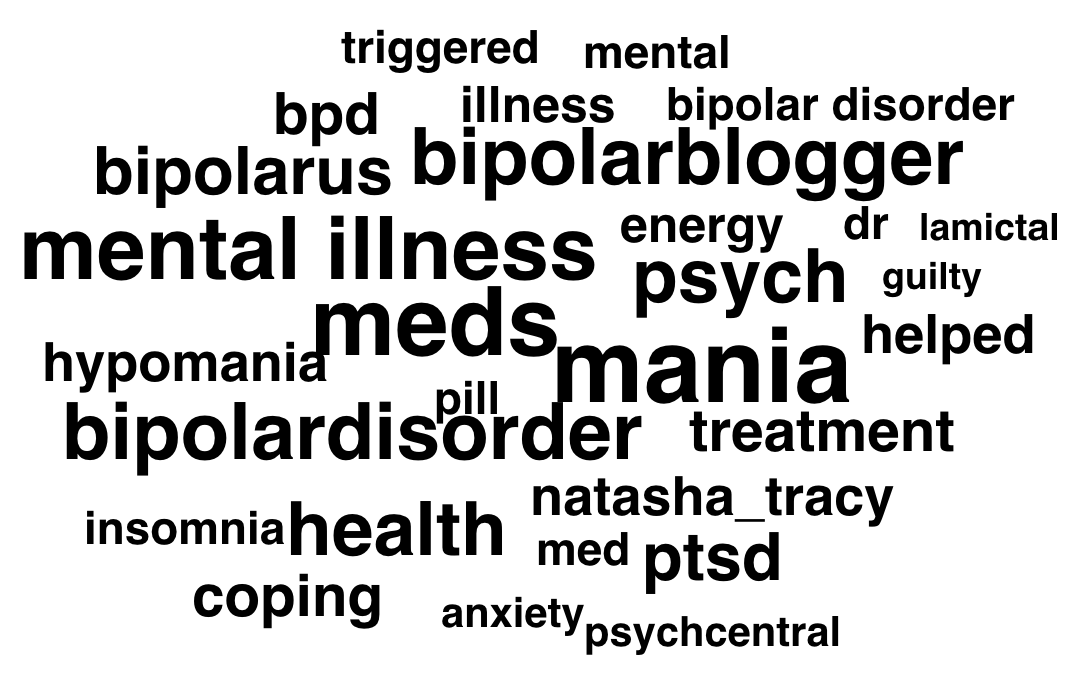}}
    \caption{SC Wordcloud}
    \label{fig:wordcloud_gb}
\end{figure}

The difference between the datasets collected by \textit{subconscious crowdsourcing (SC)} and our approach - \textit{Time-specific subconscious crowdsourcing (TSC)} is first compared in the experiment.
Two random forest classifiers were built, corresponding to the \textit{CLF} for both datasets.
With this tree based classifier, the most frequent words for each dataset were further extracted and plotted on a word cloud as shown in Fig.~\ref{fig:wordcloud_gb}.

It can be appreciated in Fig.~\ref{fig:wordcloud_gb} that many frequent words in \textit{SC} are part of the terminology of BD symptoms, such as mania, psych, PTSD, BPD, etc.

\begin{figure}[h]
    \centerline{\includegraphics[width=0.8\linewidth]{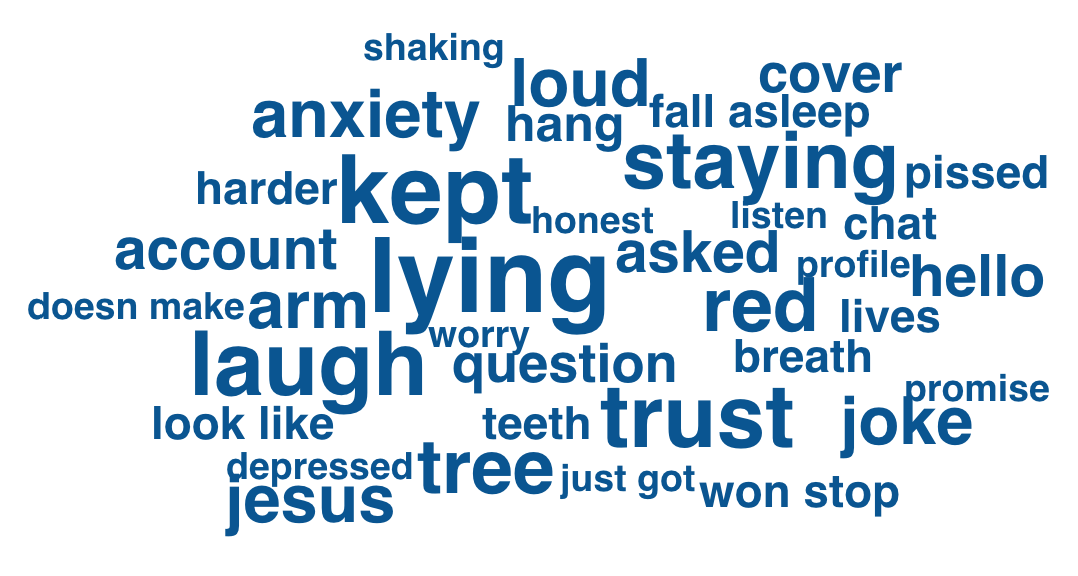}}
    \caption{TSC Wordcloud}
    \label{fig:wordcloud_eric}
\end{figure}

On the contrary, in Fig.~\ref{fig:wordcloud_eric}, the most frequent words from \textit{TSC} are closer to daily conversation phrases. 
The data collection for \textit{SC} is done from users who follow BD-related fan pages.
Consequently, among all users, those followers are more aware to their mental illness states; hence, they have higher tendency to share knowledge of BD.
It is worth pointing out that most users were not aware of their mental status before being diagnosed with BD.
The words or phrases they used were consequently less related to BD symptoms. 

As compared to \textit{SC}, our approach collected user data based on whether they mentioned being diagnosed with BD or not.
The coverage of our approach is more balanced than that of TSC, which collects more ordinary people. 
\subsubsection{\textbf{Single Category Features Performance for Psychological and Phonological Features}}

Each single category in our proposed features, \textit{BDPLF} and \textit{Phonological features}, are evaluated separately.
Among the features, \textit{social(Soc)} and \textit{mood polarity(Pol)} features performed much better at $\alpha = \{2, 3\}$ months than during other periods.
This implies that when BD is onset, the mood polarity and social behaviors become much more significant.
For \textit{emotion(Emot)} and \textit{phonological(Phon)} features, they also perform well at $\alpha = \{9, 12\}$. 
%
To further experiment on pure text features, we not only eliminated the \textit{social} features, as they fundamentally required  social interaction, but also removed the \textit{mood polarity(Pol)} category,  since it requires time interval information. 

For the \textit{age and gender(AG)} and \textit{insomnia(LT) and over-talking(TRD)} features, the major reason for having a lower performance is that there are less then 3 sub-features in this category, while the other categories have more. 
In the next section, we present further feature ensemble experiments to examine these features.

\renewcommand\tabcolsep{2pt}
\begin{table}[h]
    \centering
    \begin{tabular}{c|c|c|c|c|c} 
         \hline
         Features($\#$DIM)& 2 mths & 3 mths & 6 mths & 9 mths & 12 mths \\ [0.3ex] 
         \hline
         AG$^{(2)}$ & 0.475 & 0.503 & 0.445 & 0.434 & 0.383\\
         Pol$^{(5)}$ & \textbf{0.911} & \textbf{0.893} & 0.843 & 0.836 & 0.803\\
         Emot$^{(8)}$ &  0.893 & 0.895 & \textbf{0.908} & \textbf{0.917} & \textbf{0.896} \\
         Soc$^{(4)}$ & \textbf{0.941} & \textbf{0.913} & \textbf{0.845} & 0.834 &  0.786 \\
         LT$^{(1)}$ & 0.645 & 0.589 & 0.554 & 0.504 &  0.513 \\
         TRD$^{(1)}$ & 0.570 & 0.638 & 0.626 & 0.615 &  0.654 \\
         Phon$^{(8)}$ & 0.889 & 0.880 & 0.802 &  \textbf{0.838} & \textbf{0.821}\\[0.3ex]
         \hline
    \end{tabular}
    \caption{Average Precision of Single Feature Performance}
    \label{table:singleFeaturePrediction}
\end{table}

\subsubsection{\textbf{Bipolar Disorder Onset Prediction Model Performance}}

To compare the performance of the models proposed, we used 10-fold cross validation to measure the overall precision and recall.
To assess and interpret the performance of each classification model and each selected time period, the overall performance is summarized in Table~\ref{table:overallprecision}.
In general, all of the ensemble features performed well in the onset detection task.
The ensemble of \textit{Phonological features} and \textit{BPLF}, which is the integration of our proposed features with BD-customized features from original \textit{Pattern of Life(PLF)}, performed the best. 
This ensemble successfully captured the BD features in phonological patterns and user engagement (Soc) metrics, besides the sentiment and emotions portrayed in user tweets.
The \textit{CLM} had a lower performance based on what was discussed in the previous section.
That is, when dealing with unaware patients, the classifier is not able to recognize BD. 
Similarly, the \textit{LIWC} model only depends on psychological words and does not offer enough information in order to detect BD.

As can be seen in Table~\ref{table:overallprecision}, the models  trained on three months of user data performed the best, which also indicates that BD features are more obvious when the time period is 2 to 3 months before diagnosis.
Rapid cycling bipolar disorder, from DSM-IV, is defined as a pattern of presentation accompanied by 4 or more mood episodes in a 12-month period, with a typical course of mania or hypomania followed by depression or vice versa.
These episodes must be demarcated by a full or partial remission lasting at least 2 months or by a switch to a mood state of opposite polarity\cite{american2013diagnostic}.
These 2 to 3 month periods also match clinical observations, which indicates that our model and dataset have a high reliability.

Without taking all of the user data into consideration, the model was able to perform well even when dealing with short-time data. 
Furthermore, this 2--3 month time period is considered to be the best size for a time-frame of BD prodrome detection.
We thus set the time frame to $\alpha = 2$ \textit{months}.

\renewcommand\tabcolsep{1pt}
\begin{table}[h]
    \centering
    \begin{tabular}{c|c|c|c|c|c} 
         \hline
         Features ($\#$DIM) & 2 mths & 3 mths & 6 mths & 9 mths & 12 mths \\ [0.1ex]
         \hline
            LIWC$^{(64)}$ & 0.430 & 0.391 & 0.377 & 0.403 & 0.372\\
            CLF$^{(1000+)}$ & 0.934 & 0.927 & 0.866 & 0.840 & 0.809\\
            PLF$^{(19)}$ & 0.977 & 0.979 & 0.967 & \textbf{0.975} & \textbf{0.967}\\
            LT+PLF$^{(20)}$ & 0.976 & 0.981 & 0.965 & 0.971 & 0.960\\
            TRD+PLF$^{(20)}$ & 0.978 & 0.978 & \textbf{0.971} & \textbf{0.977} & 0.961\\
            BDPLF$^{(21)}$ & 0.981 & 0.981 & 0.968 & 0.973 & \textbf{0.970}\\
            \hline
            Phon+PLF$^{(27)}$ & 0.977 & \textbf{0.985} & 0.966 & 0.974 & 0.961\\
            Phon+LT+PLF$^{(28)}$ & 0.976 & 0.982 & 0.964 & 0.972 & 0.962\\
            Phon+TRD+PLF$^{(28)}$ & \textbf{0.983} & \textbf{0.983} & \textbf{0.971} & 0.964 & 0.958\\
            Phon+BDPLF$^{(29)}$ & \textbf{0.984} & \textbf{0.983} & 0.969 & 0.968 & 0.965\\
            \hline
            \hline
            Emot+AG$^{(10)}$ & 0.904 & 0.904 & 0.873 & 0.909 & 0.885\\
            Emot+Phon$^{(18)}$ & 0.948 & 0.947 & 0.935 & 0.938 & 0.924\\
            Emot+AG+Phon$^{(20)}$ & 0.950 & 0.950 & 0.937 & 0.940 & 0.917\\
         \hline
    \end{tabular}
    \caption{Average Precision of Model Performance}
    \label{table:overallprecision}
\end{table}

The pure text model is further evaluated in this step.
The goal of this experiment is to provide foresight for the potential patterns in human natural expressions. 
In this experiment, \textit{age and gender(AG)}, \textit{emotion(Emot)}, and \textit{phonological(Phon)} features are treated in ensemble as pure text features and we construct a predictive model through them.
The \textit{LIWC} features and \textit{CLF} are not considered as pure text features, because they already have a high dimension of features, which are $64$ and over $1000$ dimensions.

The performance of the pure text ensemble model is shown in the bottom of Table~\ref{table:overallprecision}.
It can be observed in the table that the model reaches an average precision near to 0.9 without phonological features.
With the involvement of \textit{phonological features}, the performance of the model improved by 7\% and reached around 
0.96 when the time frame was $\alpha = \{2,3\}$, which is almost competitive with the social involved model.

Moreover, the pure text model is also valuable because of the efforts reduced in processing steps and data collection with little performance loss for the BD onset detection task. 



\begin{figure}[h!]
    \centerline{\includegraphics[width=\linewidth]{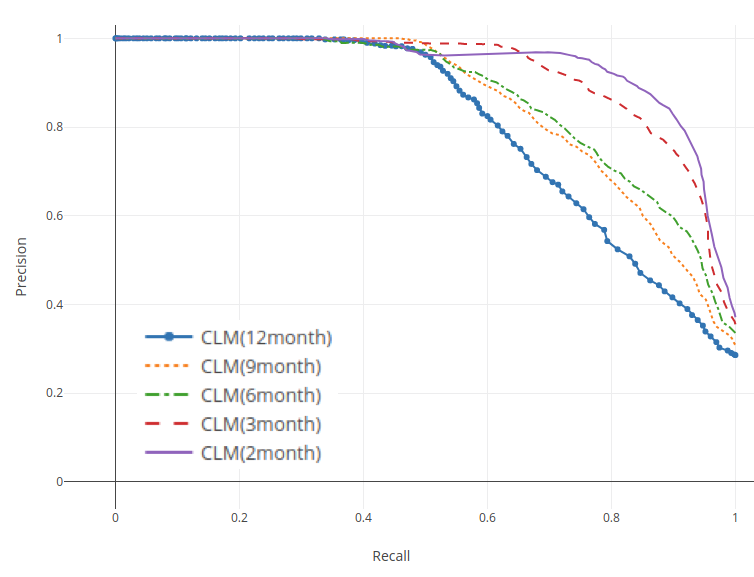}}
    \caption{Average Precision Curve (CLM)}
    \label{fig:tfidf_timely}
\end{figure}

Generally, each classifier for each time period performed better as the dataset segmentation was closer to the time of diagnosis, especially for the \textit{CLM}.
For this classifier, Figure~\ref{fig:tfidf_timely} shows that the average precision moved from $0.809$ to $0.934$ as the time frame $\alpha$ moved from $12$ months to $2$ months. 

The improved performance from time to time indicates that for the same group of regular and BD users, the word expression feature changes differently.
Words and phrases in the \textit{CLM} reflect the linguistic styles of BD and regular users, which provides useful information about how BD patients are responding with respect to their mental state.
For instance, in Figure~\ref{fig:wordcloud_eric}, besides the negative words that reveal a negative state for a BD user, we are able to infer that he/she is getting overwhelmed in the life, which is one of the symptoms of BD.

\begin{figure}[h]
    \centerline{\includegraphics[width=0.8\linewidth]{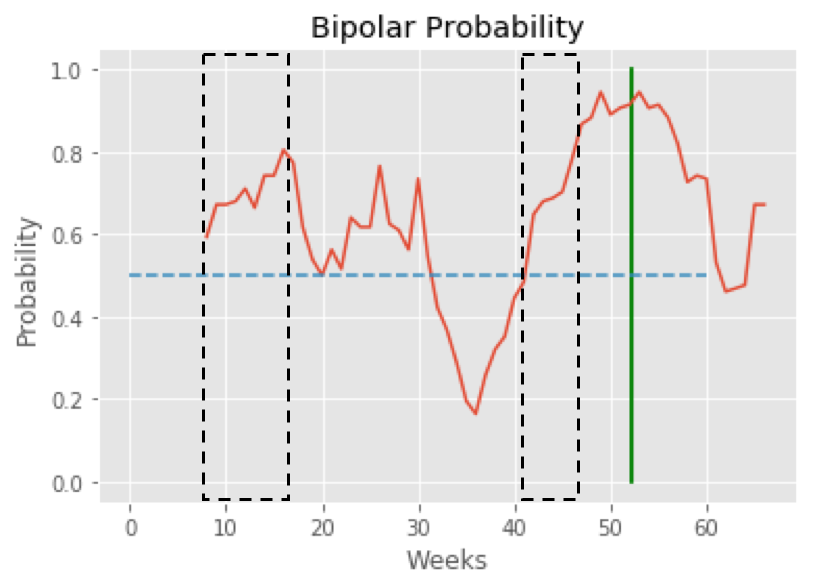}}
    \caption{Bipolar Disorder Onset Timeline}
    \label{experiment:predict_timeline}
\end{figure}

Based on the classifier, we segment user social data based on the time dimension in sliding windows approach and predict the BD onset probability for each segment. The prodrome filter approach can detect the signs of initial prodrome. For example, Figure~\ref{experiment:predict_timeline} is one of the results, the boxed periods are the most possible initial prodromal periods identified by the proposed technique.

\subsection{Psychologist Evaluation}

In this experiment, $90$ samples of user social behaviors are extracted, where each time period lasts for $2$ months. Among these $90$ samples, $1/3$ are classified as onset by \textit{BDOPM}, $1/3$ are classified as no-onset by \textit{BDOPM}, and $1/3$ are extracted from regular user behaviors and classified as no-onset. A web interface is provided to visualize these data. $8$ psychologists are involved in the blind-test experiment to assign assessments (i.e., having BD symptoms or not) to those samples. Each sample is assessed by more than three psychologists.

         

\begin{table}[h]
    \centering
    \begin{tabular}{c|c|c|c} 
         \hline
         Agreement & Onset BD & Not Onset BD & Not Onset Regular \\ [0.3ex] 
         \hline
         Majority  & 0.429 & 1.00 & 1.00  \\
         One Onset  & 0.815 & 0.857 & 0.926 \\
         \hline
    \end{tabular}
    \caption{Accuracy for Agreement Evaluation}
    \label{table:psychologist_res_agree}
\end{table}

\begin{table}[h]
    \centering
    \begin{tabular}{c|c|c} 
         \hline
         Onset BD \ &\ Not Onset BD \ &\ Not Onset Regular \\ [0.3ex] 
         \hline
         0.37 & 0.45 & 0.75 \\
         \hline
    \end{tabular}
    \caption{Agreement Rates between Psychologists}
    \label{table:psychologist_agreement}
\end{table}

Table~\ref{table:psychologist_res_agree} demonstrates that \textit{BDOPM} achieves high accuracy in most cases. 
%
%
For majority agreement, the accuracy of BD onset prediction is $0.429$ owing to the various diagnosed standards among the psychologists. Table~\ref{table:psychologist_agreement} shows that only $37\%$ of onset data among the opinions from the psychologists obtain exactly the same labels. This could be due to that the same symptom is found in many kinds of illness. 

Nearly half of the patients with BD have major depression, while roughly one-quarter to one-third of them are diagnosed as borderline personality disorder, post-traumatic stress disorder (PTSD), generalized anxiety, and social phobia were each diagnosed in roughly one-quarter to one-third~\cite{zimmerman2010psychiatric}. 
Burgess~\cite{ballard2006bipolar} also indicates there are almost $70\%$ of BD patients have been misdiagnosed more than $3$ times before receiving their correct diagnosis. 
Due to the similarity of symptoms shared by several mental disorders, chances of the majority of the psychologists agree the user as BD are low under the condition that based only on the text without a comprehensive diagnostic interview. Therefore, we give a relaxing criterion that the user is considered as BD if one of the psychologists agree that the user is BD.     

Under this relaxing criterion, the users that have never been diagnosed BD may exhibit BD-like behaviors, such as over-talk, anxiety, or struggle with the negative or angry experiences.

In summary, from different types of data, our model is able to distinguish the possible onset behavior and reach a high precision under the relaxing criterion.


\section{Conclusion}
In this research, \textit{time-specific subconscious crowdsourcing}, a novel data collection approach is proposed, which is capable to filter out not only highly self-aware but also general BD users on social media. With this approach, the time spent on diagnosis of illness could also be the key to identify the possible signs of the initial prodrome. The result shows that the \textit{pattern of life} features combined with our extending features, \textit{e.g., insomnia and over-talk feature}, provide a great performance on detecting the BD onset period. 

Furthermore, we introduce a novel \textit{phonological feature}, which focuses on the energy of words based on the evidence of phonology. By simply employing phonological feature with pure text ensemble model (i.e., the model without any social media features), the classifier can achieve more than $91\%$ precision. The proposed model also has achieve high accuracy in the onset detection task.

The time influence-\textit{CLM} indicates that the word choices of users can reflect the corresponding mental states. With this observation, it increases the possibility to perform regular assessment on people based on the simple writing.

\appendix

\bibliographystyle{ACM-Reference-Format}
\bibliography{bipolar}

\end{document}